# SNPs Filtered by Allele Frequency Improve the Prediction of Hypertension Subtypes

Yiming Li
Department of Preventive Medicine
Northwestern University
Chicago, USA
yiming.li@northwestern.edu

Sanjiv J. Shah
Division of Cardiology
Department of Medicine
Northwestern University
Chicago, USA
sanjiv.shah@northwestern.edu

Donna Arnett
College of Public Health
University of Kentucky
Lexington, USA
donna.arnett@uky.edu

Ryan Irvin
Department of Epidemiology
The University of Alabama at Birmingham
Birmingham, USA
irvinr@uab.edu

Yuan Luo
Department of Preventive Medicine
Northwestern University
Chicago, USA
yuan.luo@northwestern.edu

*Abstract*—Hypertension is the leading global cause of cardiovascular disease and premature death. Distinct hypertension subtypes may vary in their prognoses and require different treatments. An individual's risk for hypertension is determined by genetic and environmental factors as well as their interactions. In this work, we studied 911 African Americans and 1,171 European Americans in the Hypertension Genetic Epidemiology Network (HyperGEN) cohort. We built hypertension subtype classification models using both environmental variables and sets of genetic features selected based on different criteria. The fitted prediction models provided insights into the genetic landscape of hypertension subtypes, which may aid personalized diagnosis and treatment of hypertension in the future.

## I. Introduction

Hypertension is the major global cause of cardiovascular disease and premature death [1, 2]. Its subtypes may be defined by combined or isolated elevations of systolic and diastolic blood pressure, and experimental data support that these subtypes may reflect different underlying pathological mechanisms [3]. Clinical studies have also found that distinct hypertension subtypes may differ in their prognoses, and hence require different therapeutic treatments [3, 4]. Therefore, building prediction models for hypertension subtypes is relevant and useful in a clinical setting [5].

As well-established risk factors of hypertensive disorders, age and other environmental confounders like smoking were found to be differentially associated with hypertension subtypes [3]. However, previous studies have shown that a large proportion of blood pressure variability remains unexplained after controlling for the environmental factors and comorbidities (e.g., obesity, diabetes, chronic kidney disease) known to influence hypertension [6]. Additionally, previous twin and family studies have provided heritability estimates for systolic and diastolic blood pressure levels, which were between 30% and 60% [7, 8]. These studies indicate that hypertension is significantly heritable and may be associated with common genetic susceptibility [6, 9]. Hence, there exists a need for including genetic markers in hypertension subtype prediction models.

A large proportion of variation in the human genome is reflected by single-nucleotide polymorphisms (SNPs) [10]. With the emergence of large biobanks and the development of genotyping methods, genome-wide association studies (GWAS), which scan the entire genome for SNPs and other variants, were introduced. Large-scale GWAS have identified a number of loci significantly associated with hypertension [11], and it may be useful to incorporate SNPs as genetic features when predicting hypertension subtypes. Since there are many more variants than samples in GWAS data sets, selecting a subset of genetic markers as the features prior to model-fitting may help improve model performance.

Statistical and machine learning models have been successful in predicting the presence or absence of hypertension using environmental factors as well as significant SNPs from large-scale association studies [12, 13] or genetic risk scores constructed based on GWAS summary statistics [14] as the predictors. On the other hand, studies have shown that hypertension subtypes are differentially associated with demographic variables like age and sex [3, 15]. Using an ordinal logistic regression approach, previous GWAS have identified a number of loci significantly associated with different hypertension subtypes [16]. Others studied the genetics of hypertension from a different perspective and presented unsupervised approaches to defining hypertension subtypes based on genetic and/or phenotypic characteristics [17, 18]. These subtypes were next shown to have different cardiac mechanisms [18]. However, very few studies focused on predicting hypertension subtypes using both genetic and environmental variables.

In this study, we built hypertension subtype classification models using both genetic and environmental predictors. We applied two different algorithms, including multinomial logistic regression with L2 regularization, multi-layer perceptron (MLP), and ScanMap (supervised confounding aware non-negative matrix factorization for polygenic risk modelling), and compared their performances with the baseline covariates-only model as well as polygenic risk score (PRS) models. We also investigated the predictive power of various subsets of genetic features selected based on different criteria. Our fitted prediction model provided insight into the genetics of hypertension subtypes, which may be helpful for the personalized diagnosis and treatment of hypertension.

## II. MATERIALS AND METHODS

### A. Participants

Our study was based on the Hypertension Genetic Epidemiology Network (HyperGEN) data [19]. Designed for studying the genetics of hypertension, HyperGEN is part of the Family Blood Pressure Program established by the National Heart Lung and Blood Institute. Its participants included hypertensive patients, their siblings, and a random sample of age-matched normotensive controls. The details of the HyperGEN study, including its study design and sample recruitment strategy, have been well described in Williams et al. [19].

Our cohort included 911 African Americans and 1,171 European Americans. The African and European Americans were genotyped using the Affymetrix Genome-Wide SNP Array, and the GWAS data was imputed using Minimac [20] and the 1000G Phase I Integrated Release Version 3 Haplotypes (2010-11 data freeze, 2012-03-14 haplotypes) which contained haplotypes of 1092 individuals from all ethnic backgrounds with monomorphic and singleton sites excluded. Next, we performed post-imputation variant quality control by filtering out monomorphic variants as well as SNPs with low minor allele frequency (MAF $\leq$ 1%) or low imputation quality ($R^2 \leq 0.3$). After imputation and quality control, there were 7,145,793 SNPs present in both African Americans and European Americans data. To map these SNPs to genes, we annotated the genotype data based on the NCBI 37.3 gene location file using MAGMA [21]. In our data, 2,813,871 (39.38%) SNPs mapped to at least one gene, and there were 15,394 genes containing valid SNPs.

The hypertension subtype of a subject was derived from their clinic blood pressure measurements and the number of antihypertensive and diuretic medications that they were taking. All the subjects were labelled as non-hypertensive, mild hypertensive, or severe hypertensive according to the criteria from the Sixth Report of the Joint National Committee on Prevention, Detection, Evaluation, and Treatment of High Blood Pressure (JNC VI) [22]. If an individual took more than one anti-hypertensive medications, had a systolic blood pressure (SBP) over 160, or had a diastolic blood pressure (DBP) over 100, they would be defined as a severe hypertensive. A subject would also be labelled as a severe hypertensive if they took one anti-hypertensive medication and either had a greater than 140 SBP or a greater than 90 DBP. If a subject did not satisfy the severe hypertensive criteria, but took one anti-hypertensive medication, had an over 140 SBP or an over 90 DBP, they would be marked as a mild hypertensive. The other people were defined to be non-hypertensives.

In this study, we randomly split the data into 80% training (1,665 samples), 10% validation (207 samples), and 10% testing (210 samples) stratified by the hypertension subtypes of the samples. To avoid over-fitting, we ensured that people from the same family were allocated to only one of the training, validation, or testing sets.

### B. Selection of Genetic Predictors and Polygenic Scoring

Since our GWAS data is limited in sample size, we selected subsets of genetic variants as the predictive features instead of using all the SNPs in the classification model.

To begin with, it has been shown that a majority of GWAS hits are true signals and hence, may be useful for predicting the future risk of the corresponding disorders [23, 24]. Therefore, we curated a list of 609 SNPs identified to be significantly associated (p-value $< 5 \times 10^{-8}$) with hypertensive disorders by previous GWAS studies from the GWAS Catalog [25]. Among these variants, 443 were present in our GWAS data.

Secondly, potentially pathogenic variants may be more useful for the prediction of detrimental conditions like hypertension. We calculated the CADD scores [26, 27] of all the genetic variants to quantify their degree of deleteriousness. Next, we applied a cutoff of 15, which corresponded to the median of all possible canonical splice site changes and non-synonymous variants in CADD v1.0 [26, 27], to the scaled CADD scores. In this way, we filtered out 16,227 deleterious SNPs. We also converted the dosages of deleterious SNPs to hard-calls using Plink 2.0 [28]. By setting the hard-call of a dosage to be non-missing when its distance from the nearest hard-call (0, 1, or 2) was not greater than 0.1, we obtained 695 non-missing hard-calls of the identified deleterious SNPs. Additionally, we identified 3,274 genes with at least one deleterious SNP, and calculated the total dosage of deleterious SNPs in these genes.

Previous studies have found that compared with rare variants, common variants are more likely to be neutral or nearly neutral [29]. Therefore, extremely common variants may be less helpful for the risk prediction of hypertension. We performed Exome Aggregation Consortium (ExAC) [30] annotation using ANNOVAR [31], and excluded the most common SNPs by keeping only the SNPs with a less than 0.9 allele frequency in ExAC samples. 16,047 SNPs passed the filtering, among which 581 variants had a non-missing hard-call under the hard-call threshold of 0.1. There were 7,170 genes with at least one ExAC-filtered SNP, and we computed the total dosage of ExAC-filtered SNPs in these genes.

Finally, it has been shown that a PRS-based approach may outperform machine learning methods in genetic risk prediction [32]. Therefore, we also included PRS as one of the genetic predictors in our comparison. We first conducted GWAS using Plink 2.0 [28] on the training and validation samples (1,872 individuals in total) with age, sex, and the first ancestry-informative principal component as the covariates. Using the resulting GWAS summary statistics as the base dataset, polygenic scoring was next performed for all the individuals in our cohort using PRSice-2 [33]. Since PRSice-2 could only handle binary and quantitative traits [33], we coded the three hypertension subtypes as a continuous phenotype – 0 (non-hypertensive), 1 (mild hypertensive), and 2 (severe hypertensive). To avoid over-fitting, we also masked the phenotypes of testing set samples when running PRSice-2.

To summarize, we have selected eight sets of genetic predictors in total, including previous GWAS hits, deleterious SNPs (dosage or hard-call), deleterious genes, ExAC-filtered SNPs (dosage or hard-call), ExAC-filtered genes, and PRS.

### C. Hypertension Subtype Prediction

Our multiclass classification problem is to predict the hypertension subtype (non-hypertensive, mild hypertensive, or severe hypertensive) of an individual given their genetic data. To tackle this problem, we used multinomial logistic regression with L2 regularization, MLP, and ScanMap (supervised confounding aware non-negative matrix factorization for polygenic risk modelling), a novel framework capable of learning groups of genes for subsequent prediction and shown to have improved accuracy over previous methods [34]. We fitted different classification models with one set of genetic predictors and four covariates related to hypertension (age, sex, race, and whether a subject ever smoked cigarettes) as the predictors on the training set. For comparison, we also trained a baseline model predicting the hypertension subtype using only the four covariates. The model parameters were tuned using the validation set, and the predictive performance was evaluated by calculating the bootstrapped 95% confidence interval for the micro-weighted F1 score on the testing set.

To further interpret the fitted classification models, we applied the kernel SHAP (SHapley Additive exPlanations) algorithm [35] to two of the fitted logistic regression models – the covariates-only model and the best-performing model with genetic predictors. We computed the SHAP values of different features used in the models, which reflected their impact on the decision associated with each outcome class (non-hypertensive, mild hypertensive, and severe hypertensive). We also examined whether the most important genetic features (SNPs and genes) in the best-performing model were found to be associated with hypertension in previous research.

III. RESULTS

After quality control, our data contained 2,082 individuals and 7,145,793 SNPs in total. In our cohort, 342 (37.54%), 279 (30.63%), and 290 (31.83%) of the African Americans were non-hypertensive, mild hypertensive, and severe hypertensive, respectively. On the other hand, 569 (48.59%), 336 (28.69%), and 266 (22.72%) of the European Americans had no, mild, and severe hypertension, respectively.

Table I demonstrated the predictive performances of models using different types of genetic predictors and algorithms. It is well-known that hypertension is strongly correlated with age [36]. Indeed, in our results, the baseline model with age, sex, race, and smoking status as the predictors had a testing F1 of 0.6059, and many models with genetic predictors did not outperform the baseline model, possibly due to our limited sample size and the strong effect of age. However, when including the hard-calls of ExAC-filtered SNPs in the model, the testing F1 was greatly improved for ScanMap (0.6346), which was also significantly better than the best-performing PRS model (using MLP, 0.6188). This indicated that filtering out extremely common SNPs and quantifying the SNPs using hard-calls instead of dosages is helpful for the prediction of hypertension subtypes, and potentially, other disorders as well.

In the fitted covariates only model, age and race were the most important predictors for all three classes (Fig. 1A). In our results, hypertension was associated with an increased age and being African American (Fig. 1A). Race was the leading factor for predicting whether a person had severe hypertension, whereas age was the most important feature for making inference about the other two classes (non-hypertensive and mild hypertensive). In addition, we found that current or prior cigarette smoking was associated with hypertension (Fig. 1A). It was interesting that although males were more likely to be non-hypertensive than females in our data, they were also more prone to developing severe hypertension (Fig. 1A). Nevertheless, the effect of gender was quite small in our fitted model, and further studies are needed to determine whether significant gender differences exist in mild versus severe hypertensives.

Our best-performing model used the hard-calls of ExAC-filtered SNPs as the genetic predictors, and the top ten predictors associated with each hypertension status in this model were all genetic variants (Fig. 1B). Table II presented the variable importance of these SNPs (quantified by absolute SHAP values), what genes they belong to (if any), and whether they were among the top ten features associated with each class. We found that rs11079007, an intronic SNP in the gene HAP1, was the most important predictor across all three classes. HAP1 is related to neurodegeneration pathways and Huntington's disease [37], and previous studies have found hypertension to be associated with an earlier onset age of Huntington's disease as well as neurodegeneration in mouse models [38-40]. Experimental studies need to be conducted to examine whether, or how often, HAP1 influences the neurodegeneration process through a hypertensive state. The intronic variant rs1996323 was also found to be among the top ten most important predictors for all three classes. It belongs to MERTK, a protein-coding gene expressed by cardiac and arterial resident macrophages which play a protective role in the heart and blood vessels [41]. This finding may support the strategy to pharmacologically target monocytes when treating arterial hypertension, as proposed by Wenzel [41]. It is also worth noting that in our results, the most important predictors for non-hypertensive and mild hypertensive had a relatively large intersection, whereas the top predictors for severe hypertension rarely overlapped with the other two sets. This further confirmed the genetic heterogeneity of hypertension, and future studies are needed to identify and compare the genetic factors uniquely contributing to different hypertension subtypes. Such studies may help us better understand why there exists a marked heterogeneity in the patients' responses to hypertensive drugs and enable more personalized therapy for hypertension.

IV. DISCUSSION

In this study, we performed hypertension subtype prediction using both genetic and environmental predictors. In our results, ScanMap outperformed multinomial logistic regression with L2 regularization and MLP, and the best-performing model used age, race, sex, smoking, and the hard-calls of ExAC-filtered SNPs as the predictors. The fact that ExAC-filtered SNPs improved the prediction performance the most was supportive of the previous findings that common variants are more likely to be neutral [29]. Filtering out extremely common SNPs prior to model-fitting and thresholding the SNPs' imputed dosages to hard-calls may help improve the performance of future disease prediction models using GWAS variants.

It is interesting that the model using previous GWAS signals for hypertensive disorders did not outperform the ExAC-filtered SNPs model. As the curation process may be incomplete and outdated, a literature mining approach may lead to a more comprehensive list of mutational candidates [42]. Most of the studies in our curated list from the GWAS catalog were treating hypertension as a binary outcome. Therefore, the significant SNPs identified through these studies may be more helpful in predicting the presence of hypertension than the degree of hypertension or hypertension subtypes. Future studies may assess if we can improve the performance of hypertension subtype prediction models by using only the GWAS SNPs significantly associated with the severity of hypertension as the genetic predictors.

Our results demonstrated the ScanMap model with ExAC-filtered SNPs could predict hypertension subtypes better than PRS models. The reason for this may be twofold. Firstly, since the PRS is a sum of trait-associated alleles weighted by their effect sizes on a considered phenotype [43], it could only capture the linear additive effects of the underlying genetic factors. Therefore, machine learning methods could outperform traditional PRS models when the nonlinear effects and interactions of genetic markers highly contribute to the variance of the studied trait. Secondly, our PRS model was built using the summary statistics of a GWAS on a limited sample size. Hence, its predictive performance may be harmed by the GWA study's lack of statistical power to detect genetic effects. Further studies to investigate functional genetic pathways [44, 45] on a larger cohort are needed to better compare the performances of PRS and machine learning models for the prediction of hypertension subtypes.

Hypertension is rarely present in isolation, and is often associated with comorbidities like obesity, diabetes, and chronic kidney disease [46]. Hence, future prediction models may benefit from examining the correlation structure among hypertension and its comorbidities and including these disorders as additional risk factors for hypertension.

In our analysis, race was found to be an important predictor of hypertension subtype. The presence of race/ethnic differences in the genetics of hypertension has been supported by previous studies [47]. Nevertheless, the loci showing substantial ethnic differences is limited in number, and the current GWAS scans were strongly biased towards populations of European ancestry [47, 48]. To examine whether the ethnic differences in the effect of genetic markers on hypertension risk are indeed genuine, further analyses of the HyperGEN data may be conducted on African Americans and European Americans separately. Additionally, if a genetic variant is found to be constantly significant in models fitted on data from different races, we can be more confident in the broader relevance of the discovery. It will be promising to prioritize these markers in future transethnic studies, which will help boost genomic research in underrepresented populations of people who are not of European descent, or of mixed ancestry [49].

One limitation of this study was the lack of replication of our findings in a second, independent cohort. Another limitation of this study was the relatively limited sample size of our studied population. A larger GWAS cohort will potentially increase the predictive performance of our models and enable the usage of more complicated machine learning methods that require a substantial amount of data for model-fitting. Approaches that integrate multi-modal data (for example, omics and phenotype data) could also utilize the existing samples more efficiently by triangulating multiple evidences [50]. On the other hand, with a larger data set, we may stratify the cohort into several age bands and fit individual models for each age group. In our results, age was usually a very important predictor, and its effects sometimes overshadowed the other predictors. By building models specific to a certain age group, we may better examine how the effects of genetic and environmental factors on the risk of hypertension vary across the age spectrum, which will be useful for making clinical intervention decisions tailored to individual patient profiles.

TABLE I. TESTING SET PERFORMANCE OF HYPERTENSION SUBTYPE CLASSIFICATION ON THE HYPERGEN DATASET

| Model | Number of genetic predictors | Algorithm | Testing F1 (95% CI) |
|---|---|---|---|
| Covariates only | None | LR | 0.6059 (0.6006, 0.6112) |
| | | MLP | 0.6036 (0.5985, 0.6087) |
| Polygenic score | 1 (polygenic score) | LR | 0.5919 (0.5868, 0.5969) |
| | | MLP | 0.6188 (0.6139, 0.6237) |
| GWAS hits | 443 | LR | 0.5694 (0.5644, 0.5744) |
| | | MLP | 0.5717 (0.5666, 0.5769) |
| | | ScanMap | 0.5809 (0.5757, 0.5861) |
| Deleterious SNPs (dosage) | 16,227 | LR | 0.5581 (0.5529, 0.5633) |
| | | MLP | 0.5383 (0.5332, 0.5434) |
| | | ScanMap | 0.5749 (0.5696, 0.5802) |
| Deleterious SNPs (hard-call) | 695 | LR | 0.5590 (0.5539, 0.5641) |
| | | MLP | 0.5010 (0.4959, 0.5061) |
| | | ScanMap | 0.5897 (0.5846, 0.5948) |
| Deleterious genes | 3,274 | LR | 0.5678 (0.5625, 0.5731) |
| | | MLP | 0.5119 (0.5069, 0.5170) |
| | | ScanMap | 0.5991 (0.5941, 0.6041) |
| ExAC-filtered SNPs (dosage) | 16,047 | LR | 0.5986 (0.5936, 0.6036) |
| | | MLP | 0.6187 (0.6135, 0.6239) |
| | | ScanMap | 0.5910 (0.5860, 0.5960) |

| Model | Number of genetic predictors | Algorithm | Testing F1 (95% CI) |
|---|---|---|---|
| ExAC-filtered SNPs (hard-call) | 581 | LR | 0.6178 (0.6127, 0.6229) |
| | | MLP | 0.5950 (0.5899, 0.6001) |
| | | ScanMap | **0.6346 (0.6296, 0.6396)** |
| ExAC-filtered genes | 7,170 | LR | 0.5297 (0.5244, 0.5350) |
| | | MLP | 0.5568 (0.5516, 0.5620) |
| | | ScanMap | 0.5634 (0.5582, 0.5686) |

a. CI: confidence interval; GWAS: genome-wide association studies; SNP: single nucleotide polymorphism; ExAC: Exome Aggregation Consortium; LR: multinomial logistic regression; MLP: multi-layer perceptron; ScanMap: supervised confounding aware non-negative matrix factorization for polygenic risk modelling.

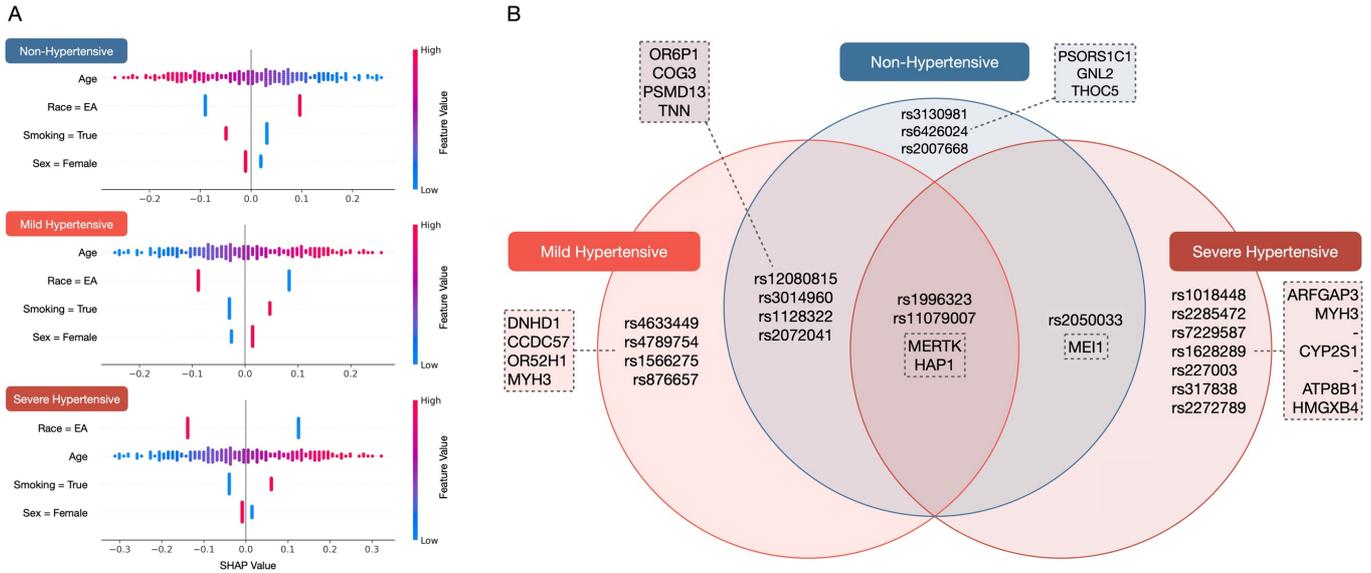

Fig. 1. The importance of features in two fitted multinomial logistic regression models for hypertension subtype prediction. (A) SHAP plot of the predictors in the covariates only model. Each subplot ranked the impact of different features on the decision associated with one hypertension status in descending order. Each point in the graph corresponded to a sample in the testing set, whose horizontal position and color were determined by the corresponding SHAP value and impact direction (red: positive; blue: negative), respectively. (B) A schematic representation of the overlap between the top ten predictors associated with each hypertension status in the ExAC-filtered SNPs (hard-call) model. The corresponding genes, if any, were also listed in the dashed rectangles. EA: European American; SHAP: SHapley Additive exPlanations; ExAC: Exome Aggregation Consortium; SNP: single-nucleotide polymorphism.

TABLE II. VARIABLE IMPORTANCE OF THE TOP TEN PREDICTORS ASSOCIATED WITH EACH HYPERTENSION SUBTYPE IN THE BEST-PERFORMING MODEL

| Variant | Gene | Non-hypertensive | | Mild hypertensive | | Severe hypertensive | |
|---|---|---|---|---|---|---|---|
| | | *Top ten* | *Importance* | *Top ten* | *Importance* | *Top ten* | *Importance* |
| rs11079007 | HAP1 | √ | 0.2160 | √ | 0.2018 | √ | 0.2551 |
| rs1996323 | MERTK | √ | 0.1376 | √ | 0.1325 | √ | 0.1493 |
| rs2072041 | TNN | √ | 0.1740 | √ | 0.1930 | | 0.1202 |
| rs1128322 | PSMD13 | √ | 0.1343 | √ | 0.1583 | | 0.0730 |
| rs3014960 | COG3 | √ | 0.1334 | √ | 0.1516 | | 0.0847 |
| rs12080815 | OR6P1 | √ | 0.1220 | √ | 0.1463 | | 0.0550 |
| rs2050033 | MEI1 | √ | 0.1320 | | 0.1215 | √ | 0.1593 |
| rs2007668 | THOC5 | √ | 0.1260 | | 0.1280 | | 0.1192 |
| rs6426024 | GNL2 | √ | 0.1247 | | 0.1190 | | 0.1388 |
| rs3130981 | PSORS1C1 | √ | 0.1210 | | 0.1125 | | 0.1414 |
| rs876657 | MYH3 | | 0.1189 | √ | 0.1391 | | 0.0665 |
| rs1566275 | OR52H1 | | 0.1179 | √ | 0.1374 | | 0.0677 |

| Variant | Gene | Non-hypertensive | | Mild hypertensive | | Severe hypertensive | |
|---|---|---|---|---|---|---|---|
| | | *Top ten* | *Importance* | *Top ten* | *Importance* | *Top ten* | *Importance* |
| rs4789754 | CCDC57 | | 0.1139 | √ | 0.1366 | | 0.0550 |
| rs4633449 | DNHD1 | | 0.1156 | √ | 0.1322 | | 0.0720 |
| rs2272789 | HMGXB4 | | 0.1075 | | 0.0659 | √ | 0.2124 |
| rs317838 | ATP8B1 | | 0.1175 | | 0.0900 | √ | 0.1906 |
| rs227003 | - | | 0.1005 | | 0.0688 | √ | 0.1786 |
| rs1628289 | CYP2S1 | | 0.0438 | | 0.0076 | √ | 0.1680 |
| rs7229587 | - | | 0.1187 | | 0.1005 | √ | 0.1654 |
| rs2285472 | MYH3 | | 0.0907 | | 0.0678 | √ | 0.1523 |
| rs1018448 | ARFGAP3 | | 0.1050 | | 0.0868 | √ | 0.1501 |

a. The variable importance was quantified by the absolute SHapley Additive exPlanations (SHAP) value.

## V. CONCLUSIONS

In this work, we fitted hypertension subtype prediction models using both environmental factors and sets of genetic features selected based on different criteria. In our study, ScanMap outperformed PRS models, multinomial logistic regression with L2 regularization and MLP, and the best-performing model used age, race, sex, smoking, and the hard-calls of ExAC-filtered SNPs as the predictors. We found that age and race were the most important predictors for all three classes (non-hypertensive, mild hypertensive, and severe hypertensive) – hypertension was associated with an increased age and African American race/ethnicity. Additionally, by examining the feature importance of different genetic variants in our best-performing model, we identified SNPs and genes useful for predicting the hypertension status of different individuals. This helped us gain insight into the genetics of hypertension subtypes, which may aid personalized diagnosis and treatment of hypertension in the future.

## ACKNOWLEDGMENT

*This project was partially supported by grants from the National Institutes of Health [U01TR003528, R01LM013337, R01HL107577, and R01HL55673].*

## REFERENCES

[1] K. T. Mills, A. Stefanescu, and J. He, "The global epidemiology of hypertension," *Nature Reviews Nephrology,* vol. 16, no. 4, pp. 223-237, 2020.
[2] V. Perkovic, R. Huxley, Y. Wu, D. Prabhakaran, and S. MacMahon, "The burden of blood pressure-related disease: a neglected priority for global health," *Hypertension,* vol. 50, no. 6, pp. 991-997, 2007.
[3] P. Verdecchia and F. Angeli, "Natural history of hypertension subtypes," ed: Am Heart Assoc, 2005.
[4] S. S. Franklin, "New interpretations of blood pressure: the importance of pulse pressure," in *Hypertension*: Elsevier, 2005, pp. 230-236.
[5] J. B. Echouffo-Tcheugui, G. D. Batty, M. Kivimäki, and A. P. Kengne, "Risk models to predict hypertension: a systematic review," *PloS one,* vol. 8, no. 7, p. e67370, 2013.
[6] G. B. Ehret and M. J. Caulfield, "Genes for blood pressure: an opportunity to understand hypertension," *European heart journal,* vol. 34, no. 13, pp. 951-961, 2013.
[7] R. Fuentes, I. Notkola, S. Shemeikka, J. Tuomilehto, and A. Nissinen, "Familial aggregation of blood pressure: a population-based family study in eastern Finland," *Journal of human hypertension,* vol. 14, no. 7, pp. 441-445, 2000.
[8] W. Miall and P. Oldham, "The hereditary factor in arterial blood-pressure," *British medical journal,* vol. 1, no. 5323, p. 75, 1963.
[9] T. Wu *et al.*, "Genetic and environmental influences on blood pressure and body mass index in Han Chinese: a twin study," *Hypertension Research,* vol. 34, no. 2, pp. 173-179, 2011.
[10] R. Sachidanandam *et al.*, "A map of human genome sequence variation containing 1.42 million single nucleotide polymorphisms," *Nature,* vol. 409, no. 6822, pp. 928-934, 2001.
[11] G. B. Ehret, "Genome-wide association studies: contribution of genomics to understanding blood pressure and essential hypertension," *Current hypertension reports,* vol. 12, no. 1, pp. 17-25, 2010.
[12] H. Izawa, Y. Yamada, T. Okada, M. Tanaka, H. Hirayama, and M. Yokota, "Prediction of genetic risk for hypertension," *Hypertension,* vol. 41, no. 5, pp. 1035-1040, 2003.
[13] C. Li *et al.*, "A prediction model of essential hypertension based on genetic and environmental risk factors in Northern Han Chinese," *International journal of medical sciences,* vol. 16, no. 6, p. 793, 2019.
[14] A. S. Havulinna *et al.*, "A blood pressure genetic risk score is a significant predictor of incident cardiovascular events in 32 669 individuals," *Hypertension,* vol. 61, no. 5, pp. 987-994, 2013.
[15] A. M. Adeoye, A. Adebiyi, B. O. Tayo, B. L. Salako, A. Ogunniyi, and R. S. Cooper, "Hypertension subtypes among hypertensive patients in Ibadan," *International journal of hypertension,* vol. 2014, 2014.
[16] C. A. German, J. S. Sinsheimer, Y. C. Klimentidis, H. Zhou, and J. J. Zhou, "Ordered multinomial regression for genetic association analysis of ordinal phenotypes at Biobank scale," *Genetic epidemiology,* vol. 44, no. 3, pp. 248-260, 2020.
[17] Y. Luo *et al.*, "Integrating hypertension phenotype and genotype with hybrid non-negative matrix factorization," in *Machine Learning for Healthcare Conference*, 2018: PMLR, pp. 102-118.
[18] Y. Ma, H. Jiang, S. J. Shah, D. Arnett, M. R. Irvin, and Y. Luo, "Genetic-Based Hypertension Subtype Identification Using Informative SNPs," *Genes,* vol. 11, no. 11, p. 1265, 2020.
[19] R. R. Williams *et al.*, "NHLBI family blood pressure program: methodology and recruitment in the HyperGEN network," *Annals of epidemiology,* vol. 10, no. 6, pp. 389-400, 2000.
[20] B. Howie, C. Fuchsberger, M. Stephens, J. Marchini, and G. R. Abecasis, "Fast and accurate genotype imputation in genome-wide association studies through pre-phasing," *Nature genetics,* vol. 44, no. 8, pp. 955-959, 2012.
[21] C. A. de Leeuw, J. M. Mooij, T. Heskes, and D. Posthuma, "MAGMA: generalized gene-set analysis of GWAS data," *PLoS computational biology,* vol. 11, no. 4, p. e1004219, 2015.
[22] Joint National Committee on Prevention, Detection, Evaluation, and Treatment of High Blood Pressure, "The 6th Report of the Joint National Committee on prevention, detection, evaluation, and treatment of high blood pressure," *Arch intern med,* vol. 157, pp. 2413-2446, 1997.
[23] V. Tam, N. Patel, M. Turcotte, Y. Bossé, G. Paré, and D. Meyre, "Benefits and limitations of genome-wide association studies," *Nature Reviews Genetics,* vol. 20, no. 8, pp. 467-484, 2019.
[24] P. M. Visscher, M. A. Brown, M. I. McCarthy, and J. Yang, "Five years of GWAS discovery," *The American Journal of Human Genetics,* vol. 90, no. 1, pp. 7-24, 2012.
[25] A. Buniello *et al.*, "The NHGRI-EBI GWAS Catalog of published genome-wide association studies, targeted arrays and summary

statistics 2019," *Nucleic acids research,* vol. 47, no. D1, pp. D1005-D1012, 2019.

[26] M. Kircher, D. M. Witten, P. Jain, B. J. O'roak, G. M. Cooper, and J. Shendure, "A general framework for estimating the relative pathogenicity of human genetic variants," *Nature genetics,* vol. 46, no. 3, pp. 310-315, 2014.

[27] P. Rentzsch, D. Witten, G. M. Cooper, J. Shendure, and M. Kircher, "CADD: predicting the deleteriousness of variants throughout the human genome," *Nucleic acids research,* vol. 47, no. D1, pp. D886-D894, 2019.

[28] C. C. Chang, C. C. Chow, L. C. Tellier, S. Vattikuti, S. M. Purcell, and J. J. Lee, "Second-generation PLINK: rising to the challenge of larger and richer datasets," *Gigascience,* vol. 4, no. 1, pp. s13742-015-0047-8, 2015.

[29] W. J. Ewens, *Mathematical population genetics: theoretical introduction*. Springer, 2004.

[30] M. Lek *et al.*, "Analysis of protein-coding genetic variation in 60,706 humans," *Nature,* vol. 536, no. 7616, pp. 285-291, 2016.

[31] K. Wang, M. Li, and H. Hakonarson, "ANNOVAR: functional annotation of genetic variants from high-throughput sequencing data," *Nucleic acids research,* vol. 38, no. 16, pp. e164-e164, 2010.

[32] D. Gola, J. Erdmann, B. Müller‐Myhsok, H. Schunkert, and I. R. König, "Polygenic risk scores outperform machine learning methods in predicting coronary artery disease status," *Genetic epidemiology,* vol. 44, no. 2, pp. 125-138, 2020.

[33] S. W. Choi and P. F. O'Reilly, "PRSice-2: Polygenic Risk Score software for biobank-scale data," *Gigascience,* vol. 8, no. 7, p. giz082, 2019.

[34] Y. Luo and C. Mao, "ScanMap: Supervised Confounding Aware Non-negative Matrix Factorization for Polygenic Risk Modeling," in *Machine Learning for Healthcare Conference*, 2020: PMLR, pp. 27-45.

[35] S. M. Lundberg and S.-I. Lee, "A unified approach to interpreting model predictions," in *Proceedings of the 31st international conference on neural information processing systems*, 2017, pp. 4768-4777.

[36] C. M. McEniery, I. B. Wilkinson, and A. P. Avolio, "Age, hypertension and arterial function," *Clinical and Experimental Pharmacology and Physiology,* vol. 34, no. 7, pp. 665-671, 2007.

[37] L. L.-y. Wu and X.-F. Zhou, "Huntingtin associated protein 1 and its functions," *Cell adhesion & migration,* vol. 3, no. 1, pp. 71-76, 2009.

[38] A. Kruyer, N. Soplop, S. Strickland, and E. H. Norris, "Chronic hypertension leads to neurodegeneration in the TgSwDI mouse model of Alzheimer's disease," *Hypertension,* vol. 66, no. 1, pp. 175-182, 2015.

[39] J. L. Schultz, L. A. Harshman, D. R. Langbehn, and P. C. Nopoulos, "Hypertension is associated with an earlier age of onset of huntington's disease," *Movement Disorders,* vol. 35, no. 9, pp. 1558-1564, 2020.

[40] J. J. Steventon, A. E. Rosser, E. Hart, and K. Murphy, "Hypertension, antihypertensive use and the delayed‐onset of Huntington's disease," *Movement Disorders,* vol. 35, no. 6, pp. 937-946, 2020.

[41] P. Wenzel, "Monocytes as immune targets in arterial hypertension," *British journal of pharmacology,* vol. 176, no. 12, pp. 1966-1977, 2019.

[42] Y. Luo, G. Riedlinger, and P. Szolovits, "Text mining in cancer gene and pathway prioritization," *Cancer informatics,* vol. 13, p. CIN. S13874, 2014.

[43] F. Dudbridge, "Power and predictive accuracy of polygenic risk scores," *PLoS genetics,* vol. 9, no. 3, p. e1003348, 2013.

[44] Y. Luo and C. Mao, "PANTHER: Pathway Augmented Nonnegative Tensor factorization for HighER-order feature learning," *arXiv preprint arXiv:2012.08580,* 2020.

[45] Z. Zeng, A. H. Vo, C. Mao, S. E. Clare, S. A. Khan, and Y. Luo, "Cancer classification and pathway discovery using non-negative matrix factorization," *Journal of biomedical informatics,* vol. 96, p. 103247, 2019.

[46] H. O. Ventura and C. J. Lavie, "Impact of comorbidities in hypertension," *Current opinion in cardiology,* vol. 31, no. 4, pp. 374-375, 2016.

[47] N. Kato, "Ethnic differences in genetic predisposition to hypertension," *Hypertension Research,* vol. 35, no. 6, pp. 574-581, 2012.

[48] M. I. McCarthy *et al.*, "Genome-wide association studies for complex traits: consensus, uncertainty and challenges," *Nature reviews genetics,* vol. 9, no. 5, pp. 356-369, 2008.

[49] A. B. Popejoy and S. M. Fullerton, "Genomics is failing on diversity," *Nature News,* vol. 538, no. 7624, p. 161, 2016.

[50] Y. Luo *et al.*, "A multidimensional precision medicine approach identifies an autism subtype characterized by dyslipidemia," *Nature Medicine,* vol. 26, no. 9, pp. 1375-1379, 2020.